\documentstyle[a4,epsf,12pt]{article}
\makeatletter
\ifcase\@ptsize
  \font\tenmsy=msbm10
  \font\sevenmsy=msbm7
  \font\fivemsy=msbm5
  \font\tenrm=cmr10
  \font\egtrm=cmr8
\or
  \font\tenmsy=msbm10 scaled \magstephalf
  \font\sevenmsy=msbm8
  \font\fivemsy=msbm6
  \font\tenrm=cmr10 
  \font\egtrm=cmr8
\or
  \font\tenmsy=msbm10 scaled \magstep1
  \font\sevenmsy=msbm8
  \font\fivemsy=msbm6
  \font\tenrm=cmr10
  \font\egtrm=cmr8
\fi
\newfam\msyfam
\textfont\msyfam=\tenmsy  
\scriptfont\msyfam=\sevenmsy
\scriptscriptfont\msyfam=\fivemsy
\def\Bbb{\ifmmode\let\next\Bbb@\else
  \def\next{\errmessage{Use \string\Bbb\space only in math mode}}\fi\next}
\def\Bbb@#1{{\Bbb@@{#1}}}
\def\Bbb@@#1{\fam\msyfam#1}
\newfam\euffam
\font\sixeuf=eufm6
\font\eighteuf=eufm8
\font\twelveeuf=eufm10 scaled\magstep1
\textfont\euffam=\twelveeuf
\scriptfont\euffam=\eighteuf
\scriptscriptfont\euffam=\sixeuf

\makeatother

\newcommand{\BQ}{{\Bbb{Q}}}
\newcommand{\BR}{{\Bbb{R}}}
\newcommand{\BZ}{{\Bbb{Z}}}
\newcommand{\Bid}{1\!{\rm l}}
\def\pn{\par\noindent}

\def\w{{\cal W}}

\def\eps{\varepsilon}
\def\lam{\lambda}
\def\be{\begin{equation}}
\def\ee{\end{equation}}
\def\ba{\begin{array}}
\def\ea{\end{array}}
\def\bea{\begin{eqnarray}}
\def\eea{\end{eqnarray}}
\def\bl{\begin{list}{}{}}

\def\ds{\displaystyle}

\newcommand{\reseteqn}{\setcounter{equation}{0}}
\newcommand{\mysection}{\reseteqn\section}
\renewcommand{\thefootnote}{\fnsymbol{footnote}}
\newtheorem{claim}{$\bullet$}
\if@twoside
\else
\fi
\begin{document}
\pagestyle{empty}
\begin{raggedleft}
BONN-HE-93-43\\
hep-th/9312097\\
December 1993\\
\end{raggedleft}
$\phantom{x}$\vskip 1cm\par
{\huge \begin{center}
Curiosities at $c$-effective $=1$
\end{center}}\par
\vfill
\begin{center}
$\phantom{X}$\\
{\Large Michael A.I.\ Flohr\footnote[1]{Supported by the Deutsche
Forschungsgemeinschaft\\
$\phantom{xxll}$email: unp055@ibm.rhrz.uni-bonn.de}}\\[5ex]
{\em Physikalisches Institut der Universit\"at Bonn\\
Nussallee 12\\
D-53115 Bonn\\ Germany\\
}
\end{center}\par
\vfill
\begin{abstract}
\noindent 
The moduli space of all rational conformal quantum field theories with
effective central charge $c_{{\em eff}} = 1$ is considered. Whereas the
space of unitary theories essentially forms a manifold, the non unitary 
ones form a fractal which lies dense in the parameter plane. Moreover, 
the points of this set are shown to be in one-to-one correspondence with 
the elements of the modular group for which an action on this set is defined.
\end{abstract}
\vfill\vfill
\newpage
\setcounter{page}{1}
\renewcommand{\thefootnote}{\arabic{footnote}}
\pagestyle{plain}
\mysection{Introduction}
Since the invention of two dimensional conformal quantum field theory by
BPZ \cite{BPZ}, one of the great outstanding problems is the
classification of all rational conformal field theories (RCFTs). BPZ
have found the discrete series of minimal models, and later all unitary
RCFTs with $c = 1$ were be classified \cite{DVV,Gin,Kir}. The study
of $\w$-algebras proved to be fruitful for the search of new rational 
models \cite{WA1,WA2}. In particular, $\w$-algebras were found which exist 
only for some special values of the central charge. In a recent work 
\cite{Flo} we have completed the classification of all theories
with $c_{{\em eff}} = c - 24h_{{\em min}} \leq 1$. Here $h_{{\em min}}$ 
denotes the energy of the minimal highest weight state. There we constructed
a series of new rational models (coming from certain $\w$-algebras).
These theories exist for $c = 1 - 8k, k \in \BZ$ and are $c_{{\em eff}}
= 1$ theories. 
\pn
In the following we will study the moduli space ${\cal M}$ of all theories 
with $c_{{\em eff}} = 1$. We write ${\cal M} = M\cup M'$, where $M$ denotes 
the well known space of unitary $c = 1$ theories and $M'$ denotes the space 
of non unitary $c_{{\em eff}} = 1$ theories. With ${\cal R}(X)$ we denote 
the set of RCFTs in a moduli space $X$.
\pn
Considering the moduli space of non unitary RCFTs, ${\cal R}(M')$, one 
obtains a strange result, see figure 2. In this paper we will answer several
questions which naturally arise from this picture: Is this set ${\cal R}(M')$ 
dense in $(\BR_+)^2$? Is there an explanation of the quadric curves? Most 
important, are the points of ${\cal R}(M')$ rational points in a manifold of 
generically non rational conformal field theories (CFTs), or are they isolated?
Finally, will the space of all CFTs look like that even for $c_{{\em
eff}} > 1$?
\pn
The last question cannot be answered yet. The others will be answered
here. Surprisingly, the structure of the set ${\cal R}(M')$ is highly 
``chaotic'' and forms a self similar fractal, nonetheless dense in $\BR^2$. 
Moreover, its points are in one-to-one correspondence to the elements of 
the modular group PSL$(2,\BZ)$. This correspondence suggests an
application of these theories to phenomena like the fractional quantum
Hall effect.

\mysection{Preliminaries}
In a recent work \cite{Flo} new RCFTs with effective central charge 
$c_{{\em eff}} = 1$ have been discovered and classified which possess 
extended symmetry algebras ($\w$-algebras). In the following, let 
$\eps\in\{0,1\}$ and let $\min\lam$ denote the smallest representative of
$\lam\in\BZ/m\BZ$. Then we have shown that for every $k$ there 
exist two RCFTs, one with
\be\ba{rcll}
  c & = & 1 - 24k\,, & k\in\BZ_+/2 \\
  {\ds h_{\frac{\lam}{2k+2\eps},(-)^{\eps}\frac{\lam}{2k+2\eps}}} &
      = & {\ds \left[\left(\frac{\min\lam}{2k+2\eps}\right)^2 - 1\right]k
          + \eps\left(\frac{\min\lam}{2k+2\eps}\right)^2}\,, &
                    \lam\in\BZ/(k+\eps)\BZ \\
  h_{1,1} & = & 0
\ea\ee
which has the extended chiral symmetry algebra $\w(2,3k)$, and its
$\BZ_2$ orbifold
\be\ba{rcll}
  c & = & 1 - 24k\,, & k\in\BZ_+/4 \\
  {\ds h_{\frac{\lam}{4k+4\eps},(-)^{\eps}\frac{\lam}{4k+4\eps}}} &
      = & {\ds \left[\left(\frac{\min\lam}{4k+4\eps}\right)^2 - 1\right]k
          + \eps\left(\frac{\min\lam}{4k+4\eps}\right)^2}\,, &
                    \lam\in\BZ/(4k+4\eps)\BZ \\
  h_{1,1} & = & 0 \\
  h_{2,2} & = & 3k
\ea\ee
with chiral symmetry algebra $\w(2,8k)$. Here $h_{r,s}$ denotes the
Virasoro highest weight analogous to the Virasoro highest weights 
of degenerate models to generic central charge $c = 1 - 24\alpha_0^2$
given by $h_{r,s} = \frac{1}{4}\left((r\alpha_-+s\alpha_+)^2
- (\alpha_-+\alpha_+)^2\right)$, where $\alpha_{\pm} = \alpha_0 \pm
\sqrt{1+\alpha_0^2}$. Note that in contrast to the generic degenerate
Virasoso model $s = \pm r$ and $r$ is not restricted to integers only.
For further details see \cite{Flo}. All these RCFTs indeed have
effective central charge $c_{{\em eff}} = c - 24h_{{\em min}} = 1$.
\pn
The finitely many highest weight representations are highest weight
representations with respect to the extended symmetry algebra. The
characters which are infinite sums of Virasoro characters can be
expressed in terms of Jacobi-Riemann $\Theta$-functions divided by the
usual Dedekind $\eta$-function. For example the vacuum character of the
$\w(2,3k)$ theories is given by
\bea
  \chi^{}_{{\em vac}}(\tau) & = & {\ds\sum_{n\in\BZ_+}\chi_{h_{n,n}}^{\em
                               Vir}(\tau) = q^{\frac{1-c}{24}}\sum_{n\in\BZ_+}
                               \frac{q^{h_{n,n}}-q^{h_{n,-n}}}{\eta(\tau)}} \\
                         & = & {\ds \frac{1}{2\eta(\tau)}\left(
                               \Theta_{0.k}(\tau) - \Theta_{0,k+1}(\tau)
                               \right)}\,,
\eea
where $\eta(\tau) = q^{1/24}\prod_{n=1}^{\infty}(1-q^n)$,
$\Theta_{\lam,k}(\tau) = \sum_{n\in\BZ}q^{(2kn+\lam)^2/4k}$, and
$q = e^{2\pi i\tau}$. The other characters can be obtained by the modular
transformation $S:\tau\mapsto-\frac{1}{\tau}$. Details including
$S$ and $T$ matrix and fusion rules again can be found in \cite{Flo}.
\pn
The partition function $Z(\tau,\bar{\tau})$ is diagonal in the
$\w$-characters $\chi^{}_{\lam,\eps}$ (obvious notation), if the
symmetric theory is chosen. Therefore, the $\w$-algebra is the
maximally extended symmetry algebra. In this case one has for the
$\w(2,3k)$ theories
\bea
  Z(\tau,\bar{\tau}) & = & \sum_{\lam,\eps}\left|\chi^{}_{\lam,\eps}\right|^2\\
                     & = & \frac{1}{2\eta(\tau)\eta(\bar{\tau})}\left(
                           \sum_{\lam=0}^{2k-1}\left|\Theta_{\lam,k}
                           \right|^2 + \sum_{\lam=0}^{2k}\left|
                           \Theta_{\lam,k+1}\right|^2\right) \\
                     & = & \frac{1}{2}\left(Z[k] + Z[k+1]\right)\,,
\eea
where we denote with $Z[2R^2] \equiv Z(R)$ the partition function of an
U(1) theory of S$^1$ mappings with fixed compactification radius $R$.
This notation is chosen for later convenience. The U(1) partition
function is given by
\be
  Z[k] = \frac{1}{\eta\bar{\eta}}\sum_{m,n\in\BZ}
            q^{\frac{1}{4k}(n + km)^2}\bar{q}^{\frac{1}{4k}(n-km)^2}\,.
\ee
This function is modular invariant and, moreover, possesses the well
known duality property $Z[k] = Z[\frac{1}{k}]$.
Of course, non diagonal partition functions can be obtained for
theories which are not symmetrically composed from the chiral and
antichiral part. These non diagonal theories are related to the
diagonal ones via automorphisms of the fusion rules.

\mysection{Classification}
First, let us recall some important identities for U(1) partition
functions for compactification radii which are square roots of rational
numbers: Let $k,p,q\in\BZ_+$. In that case one has
\bea
  Z[k] & = & \frac{1}{\eta\bar{\eta}}\sum_{1\leq n\leq 2k}\left|
             \Theta_{n,k}\right|^2\,,\\
  Z[\frac{p}{q}] & = & \frac{1}{\eta\bar{\eta}}\sum_{n\,{\rm mod}\,2pq}
                     \Theta_{n,pq}(\tau)\Theta_{n',pq}(\bar{\tau})\,,
\eea
where $n' = qr + ps$ mod $2pq$ if $n = qr - ps$ mod $2pq$. Then all
functions of the form
\be\label{eq:zeds}
  Z = \sum_{i}a_iZ[x_i],\ \ x_i\in\BQ
\ee
are modular invariant and correspond to finite dimensional
representations of (congruence subgroups) of PSL(2,$\BZ$). 
But only a small subset of these functions is
physically relevant and related to RCFTs. The physical requirements are
1.) that the power series of the characters in $q$ has non negative integer 
coefficients, 2.) that the first non vanishing coefficient is equal to one 
and 3.) that the vacuum character does not contain currents (in order to
be a Virasoro character). These requirements strongly restrict the possible 
linear combinations of type (\ref{eq:zeds}).
\pn
One of the main results in \cite{Flo} is the classification of all RCFTs
with effective central charge $c_{{\em eff}} = 1$. The unitary case has
been treated earlier \cite{DVV,Gin,Kir}. All unitary RCFTs with
$c=1$ fit in the A-D-E classification of finite subgroups of SU(2).
The corresponding physical relevant partition functions are all given
by linear combinations of U(1) partition functions corresponding
to certain congruence subgroups of PSL(2,$\BZ$). In \cite{Kir} the
completeness of this set of known partition functions (up to
non-congruence subgroups of PSL(2,$\BZ$)) has been proven with the help
of the Serre-Stark theorem. Nevertheless one additional partition
function appeared which was rejected there because it could not belong
to an unitary theory. It is of the form
\be\label{eq:myzed}
  Z_{{\em new}}[x,y] = \frac{1}{2}\left(Z[x] + Z[y]\right)\,,
\ee
where $x,y$ are rational numbers. We have the following classification
\cite{Flo}:
\begin{claim}
  Partition functions of type (\ref{eq:myzed}) are physically
  relevant, i.e.\ belong to theories with Virasoro vacuum characters,
  $\chi_{{\it vac}} = \frac{q^{-c/24}}{\eta(\tau)}(1 - q + \ldots)$,
  iff $x,y\in\BQ_+$ such that
  \be
    x=\frac{p}{q},\ \ \ y=\frac{p'}{q'}, \ \ \ p'q'-pq \in \{1,4\}\,.
  \ee
  Then $c = 1 - 24pq$.
\end{claim}
This matches exactly the series of $\w(2,3k)$ and $\w(2,8k)$ algebras
for $p'q'-pq = 1$ and $p'q'-pq = 4$, respectively. In the case of $x$ or $y
\not\in\BZ_+$ or, due to duality, $1/x$ or $1/y\not\in\BZ_+$
we have an automorphism of the fusion rules.
\pn
The surprising fact is that only certain pairs of rational numbers yield
RCFTs. In the following we will concentrate on the case of $\w(2,3k)$
theories, since the $\w(2,8k)$ theories can be understood as their
$\BZ_2$ orbifolds (for the case of $k\in\BZ_++\frac{1}{n}$, $n = 2,4$ see
\cite{Flo}).

\mysection{Moduli Space}
In the work \cite{Gin} the moduli space of the $c = 1$ theories 
has been considered. The well known picture of the moduli space which also
indicates the flow of marginal perturbations has to be modified in the
manner shown in figure 1 in order to accomodate the non unitary theories.
\begin{figure}[h]
\setlength{\unitlength}{0.0094in}%
\def\ss{\scriptstyle}
\begin{picture}(492,445)(78,340)
\thicklines
\put(200,520){\circle*{4}}
\put(320,520){\circle*{4}}
\put(260,520){\circle*{4}}
\put(230,520){\circle*{4}}
\put(360,520){\circle*{4}}
\put(420,520){\circle*{4}}
\put(320,536){\circle*{4}}
\put(320,551){\circle*{4}}
\put(320,570){\circle*{4}}
\put(320,585){\circle*{4}}
\put(320,615){\circle*{4}}
\put(320,645){\circle*{4}}
\put(120,740){\circle*{4}}
\put(120,720){\circle*{4}}
\put(120,700){\circle*{4}}
\put(120,580){\circle{4}}
\put( 80,520){\circle{4}}
\put( 83,481){\circle{4}}
\put(320,400){\circle{4}}
\put(320,520){\vector( 0, 1){240}}
\put(320,520){\vector( 1, 0){240}}
\put(200,520){\line( 1, 0){120}}
\multiput( 80,520)(7.74194,0.00000){16}{\line( 1, 0){  3.871}}
\multiput(320,520)(0.00000,-7.74194){16}{\line( 0,-1){  3.871}}
\put(320,400){\line( 0, 1){  5}}
\multiput(200,400)(0.00000,9.00000){41}{\makebox(0.5926,0.8889){\tenrm .}}
\put(200,520){\vector( 4,-3){240}}
\multiput(200,520)(-32.29630,24.22222){3}{\line(-4, 3){ 15.407}}
\multiput(200,520)(-33.90969,-11.30323){4}{\line(-3,-1){ 18.271}}
\put(202,521){\vector( 3, 1){300}}
\multiput(500,620)(-5.00000,-5.00000){3}{\makebox(0.5926,0.8889){\tenrm .}}
\multiput(490,610)(10.00000,-2.50000){3}{\makebox(0.5926,0.8889){\tenrm .}}
\multiput(510,605)(-5.00000,-6.66667){4}{\makebox(0.5926,0.8889){\tenrm .}}
\multiput(495,585)(8.00000,-2.00000){6}{\makebox(0.5926,0.8889){\tenrm .}}
\multiput(535,575)(-7.50000,-2.50000){5}{\makebox(0.5926,0.8889){\tenrm .}}
\multiput(505,565)(8.00000,-2.00000){6}{\makebox(0.5926,0.8889){\tenrm .}}
\multiput(545,555)(-7.50000,-2.50000){3}{\makebox(0.5926,0.8889){\tenrm .}}
\multiput(530,550)(10.00000,-2.50000){3}{\makebox(0.5926,0.8889){\tenrm .}}
\multiput(550,545)(-6.66667,-5.00000){4}{\makebox(0.5926,0.8889){\tenrm .}}
\multiput(530,530)(7.50000,-2.50000){5}{\makebox(0.5926,0.8889){\tenrm .}}
\multiput(560,520)(-10.00000,-5.00000){2}{\makebox(0.5926,0.8889){\tenrm .}}
\multiput(550,515)(5.00000,-5.00000){3}{\makebox(0.5926,0.8889){\tenrm .}}
\multiput(560,505)(-8.00000,-2.00000){6}{\makebox(0.5926,0.8889){\tenrm .}}
\multiput(520,495)(8.00000,-2.00000){6}{\makebox(0.5926,0.8889){\tenrm .}}
\multiput(560,485)(-7.50000,-5.00000){5}{\makebox(0.5926,0.8889){\tenrm .}}
\multiput(530,465)(5.00000,-10.00000){3}{\makebox(0.5926,0.8889){\tenrm .}}
\multiput(540,445)(-10.00000,0.00000){3}{\makebox(0.5926,0.8889){\tenrm .}}
\multiput(520,445)(5.00000,-5.00000){3}{\makebox(0.5926,0.8889){\tenrm .}}
\multiput(530,435)(-9.28571,0.00000){8}{\makebox(0.5926,0.8889){\tenrm .}}
\multiput(465,435)(8.33333,-4.16667){7}{\makebox(0.5926,0.8889){\tenrm .}}
\multiput(515,410)(-8.00000,-2.00000){6}{\makebox(0.5926,0.8889){\tenrm .}}
\multiput(475,400)(7.50000,-5.00000){3}{\makebox(0.5926,0.8889){\tenrm .}}
\multiput(490,390)(-9.00000,0.00000){6}{\makebox(0.5926,0.8889){\tenrm .}}
\multiput(445,390)(7.50000,-2.50000){5}{\makebox(0.5926,0.8889){\tenrm .}}
\multiput(475,380)(-10.00000,0.00000){5}{\makebox(0.5926,0.8889){\tenrm .}}
\multiput(435,380)(5.00000,-7.50000){3}{\makebox(0.5926,0.8889){\tenrm .}}
\multiput(445,365)(-8.33333,0.00000){4}{\makebox(0.5926,0.8889){\tenrm .}}
\multiput(420,365)(5.00000,-6.66667){4}{\makebox(0.5926,0.8889){\tenrm .}}
\multiput(435,345)(5.00000,-5.00000){2}{\makebox(0.5926,0.8889){\tenrm .}}
\multiput(200,520)(8.82353,2.20588){35}{\makebox(0.5926,0.8889){\tenrm .}}
\multiput(500,595)(0.00000,5.00000){2}{\makebox(0.5926,0.8889){\tenrm .}}
\multiput(200,520)(8.64865,-2.16216){38}{\makebox(0.5926,0.8889){\tenrm .}}
\multiput(520,440)(5.00000,0.00000){2}{\makebox(0.5926,0.8889){\tenrm .}}
\multiput(200,520)(8.12500,-4.06250){33}{\makebox(0.5926,0.8889){\tenrm .}}
\multiput(460,390)(0.00000,5.00000){2}{\makebox(0.5926,0.8889){\tenrm .}}
\multiput(200,520)(7.50000,-5.00000){31}{\makebox(0.5926,0.8889){\tenrm .}}
\multiput(425,370)(5.00000,-5.00000){2}{\makebox(0.5926,0.8889){\tenrm .}}
\multiput(200,520)(7.75862,-4.65517){30}{\makebox(0.5926,0.8889){\tenrm .}}
\multiput(425,385)(5.00000,0.00000){2}{\makebox(0.5926,0.8889){\tenrm .}}
\multiput(200,520)(8.33333,-3.33333){34}{\makebox(0.5926,0.8889){\tenrm .}}
\multiput(475,410)(5.00000,0.00000){2}{\makebox(0.5926,0.8889){\tenrm .}}
\multiput(200,520)(8.63636,-2.87879){34}{\makebox(0.5926,0.8889){\tenrm .}}
\multiput(485,425)(-5.00000,0.00000){2}{\makebox(0.5926,0.8889){\tenrm .}}
\multiput(200,520)(8.91892,-1.48649){38}{\makebox(0.5926,0.8889){\tenrm .}}
\multiput(200,520)(8.82353,1.47059){35}{\makebox(0.5926,0.8889){\tenrm .}}
\multiput(200,520)(8.78378,-1.75676){38}{\makebox(0.5926,0.8889){\tenrm .}}
\multiput(525,455)(10.00000,0.00000){2}{\makebox(0.5926,0.8889){\tenrm .}}
\multiput(200,520)(8.82353,1.76471){35}{\makebox(0.5926,0.8889){\tenrm .}}
\multiput(500,580)(0.00000,5.00000){2}{\makebox(0.5926,0.8889){\tenrm .}}
\multiput(400,380)(6.25000,6.25000){17}{\makebox(0.5926,0.8889){\tenrm .}}
\multiput(380,400)(6.15385,6.15385){14}{\makebox(0.5926,0.8889){\tenrm .}}
\multiput(360,420)(6.66667,6.66667){10}{\makebox(0.5926,0.8889){\tenrm .}}
\multiput(340,440)(6.66667,6.66667){7}{\makebox(0.5926,0.8889){\tenrm .}}
\multiput(320,460)(6.66667,6.66667){7}{\makebox(0.5926,0.8889){\tenrm .}}
\multiput(380,480)(6.66667,6.66667){4}{\makebox(0.5926,0.8889){\tenrm .}}
\multiput(420,480)(6.66667,6.66667){4}{\makebox(0.5926,0.8889){\tenrm .}}
\multiput(460,480)(6.66667,6.66667){4}{\makebox(0.5926,0.8889){\tenrm .}}
\multiput(500,480)(6.66667,6.66667){4}{\makebox(0.5926,0.8889){\tenrm .}}
\multiput(280,460)(6.66667,6.66667){7}{\makebox(0.5926,0.8889){\tenrm .}}
\multiput(260,480)(6.66667,6.66667){4}{\makebox(0.5926,0.8889){\tenrm .}}
\multiput(460,600)(6.66667,-6.66667){10}{\makebox(0.5926,0.8889){\tenrm .}}
\multiput(440,580)(6.66667,-6.66667){7}{\makebox(0.5926,0.8889){\tenrm .}}
\multiput(400,580)(6.66667,-6.66667){7}{\makebox(0.5926,0.8889){\tenrm .}}
\multiput(380,560)(6.66667,-6.66667){4}{\makebox(0.5926,0.8889){\tenrm .}}
\multiput(340,560)(6.66667,-6.66667){4}{\makebox(0.5926,0.8889){\tenrm .}}
\put(190,505){\makebox(0,0)[lb]{\raisebox{0pt}[0pt][0pt]{\egtrm 
$\ss 1/\sqrt{2}$}}}
\put(225,505){\makebox(0,0)[lb]{\raisebox{0pt}[0pt][0pt]{\egtrm 
$\ss\sqrt{3}/2$}}}
\put(260,505){\makebox(0,0)[lb]{\raisebox{0pt}[0pt][0pt]{\egtrm 
$\ss 1$}}}
\put(320,505){\makebox(0,0)[lb]{\raisebox{0pt}[0pt][0pt]{\egtrm 
$\ss\sqrt{2}$}}}
\put(360,505){\makebox(0,0)[lb]{\raisebox{0pt}[0pt][0pt]{\egtrm 
$\ss\sqrt{3}$}}}
\put(420,505){\makebox(0,0)[lb]{\raisebox{0pt}[0pt][0pt]{\egtrm 
$\ss 3/\sqrt{2}$}}}
\put(290,525){\makebox(0,0)[lb]{\raisebox{0pt}[0pt][0pt]{\egtrm 
$\ss 1/\sqrt{2}$}}}
\put(290,540){\makebox(0,0)[lb]{\raisebox{0pt}[0pt][0pt]{\egtrm 
$\ss\sqrt{3}/2$}}}
\put(290,570){\makebox(0,0)[lb]{\raisebox{0pt}[0pt][0pt]{\egtrm 
$\ss\sqrt{6}/2$}}}
\put(290,590){\makebox(0,0)[lb]{\raisebox{0pt}[0pt][0pt]{\egtrm 
$\ss\sqrt{2}$}}}
\put(290,615){\makebox(0,0)[lb]{\raisebox{0pt}[0pt][0pt]{\egtrm 
$\ss\sqrt{3}$}}}
\put(290,555){\makebox(0,0)[lb]{\raisebox{0pt}[0pt][0pt]{\egtrm 
$\ss 1$}}}
\put(290,645){\makebox(0,0)[lb]{\raisebox{0pt}[0pt][0pt]{\egtrm 
$\ss 3/\sqrt{2}$}}}
\put(290,740){\makebox(0,0)[lb]{\raisebox{0pt}[0pt][0pt]{\egtrm 
$\ss n/\sqrt{2}$}}}
\put(340,740){\makebox(0,0)[lb]{\raisebox{0pt}[0pt][0pt]{\egtrm 
$\ss D_{n+2}$}}}
\put(310,775){\makebox(0,0)[lb]{\raisebox{0pt}[0pt][0pt]{\egtrm 
$\ss R_{{\rm orb}}$}}}
\put(570,520){\makebox(0,0)[lb]{\raisebox{0pt}[0pt][0pt]{\egtrm 
$\ss R_{{\rm circle}}$}}}
\put(520,505){\makebox(0,0)[lb]{\raisebox{0pt}[0pt][0pt]{\egtrm 
$\ss n/\sqrt{2}$}}}
\put(520,530){\makebox(0,0)[lb]{\raisebox{0pt}[0pt][0pt]{\egtrm 
$\ss A_{2n-1}$}}}
\put(130,741){\makebox(0,0)[lb]{\raisebox{0pt}[0pt][0pt]{\egtrm 
$\ss E_6$}}}
\put(130,720){\makebox(0,0)[lb]{\raisebox{0pt}[0pt][0pt]{\egtrm 
$\ss E_7$}}}
\put(130,701){\makebox(0,0)[lb]{\raisebox{0pt}[0pt][0pt]{\egtrm 
$\ss E_8$}}}
\put(510,620){\makebox(0,0)[lb]{\raisebox{0pt}[0pt][0pt]{\egtrm 
$\ss R_1$}}}
\put(445,340){\makebox(0,0)[lb]{\raisebox{0pt}[0pt][0pt]{\egtrm 
$\ss R_2$}}}
\put(345,460){\makebox(0,0)[lb]{\raisebox{0pt}[0pt][0pt]{\egtrm 
Plane of non-unitarity}}}
\put(345,525){\makebox(0,0)[lb]{\raisebox{0pt}[0pt][0pt]{\egtrm 
Line of unitarity: $\ss R_1=R_2$}}}
\end{picture}
  \medskip\pn
  {\small Fig.\ 1. Survey of $c_{{\em eff}}=1$ models. The diagonal
  axis in the $R_1,R_2$ plane represents compactification on a circle
  S$^1$ with radius $R_{{\rm circle}}$, the vertical axis represents
  compactifications on the orbifold S$^1/\BZ_2$ with radius
  $R_{{\rm orb}}$. The dashed regions of theses lines are determined
  by the duality $R\leftrightarrow 1/2R$. The non unitary models
  lie in the cut plane $\{R_1,R_2 | R_1\neq R_2\}$.}
\end{figure}
First, we note that the moduli space is now three dimensional (up to the
three exceptional points for the theories with symmetry group E$_6$, E$_7$ 
and E$_8$ modded out). We used
the notation $Z_n \equiv Z(n/\sqrt{2}) = Z[n^2]$. The diagonal in the
$R_1,R_2$ plane is the line of unitarity. The points with $2R_{{\rm
circle}}^2$ or $2R_{{\rm orb}}^2\in\BQ_+$ belong to rational unitary $c = 1$ 
theories, the points in the $R_1,R_2$ plane which fulfill theorem 1 represent 
rational non unitary theories. 
\pn
For the unitary $c = 1$ CFTs one has well known marginal flows which
connect the RCFTs via irrational theories. This is not longer true for
the non unitary RCFTs with $c_{{\em eff}} = 1$. On the contrary we have
\begin{claim}
  There exists no marginal flows between the non unitary RCFTs with
  $c_{{\it eff}} = 1$. 
\end{claim}
This follows, since the $(1,1)$ field $v$ which could generate the
marginal flow, has non vanishing self coupling $C_{vv}^v$. In fact,
the field $v$ belongs to the $\w$-conformal family\footnote[1]{the family 
defined with respect to the whole chiral symmetry algebra, not just with 
respect to the Virasoro field.} to the lowest Virasoro eigenvalue, 
$(h_{{\em min}},h_{{\em min}})$, whose self fusion coefficient
$N_{{\em min},{\em min}}^{{\em min}}$ does not vanish \cite{Flo}.
It has been shown \cite{MS} that, for theories with maximally extended symmetry
algebra as in our case, this implies the non vanishing of the self coupling of
the primary fields contained in the considered $\w$-conformal family,
since their self coupling constants must be proportional to
$C_{{\em min},{\em min}}^{{\em min}}$.
\pn 
Let us now take a closer look at the $R_1,R_2$ plane. Rational non
unitary theories are represented by a certain subset ${\cal R}(M')\subset
\{x,y\in\BQ_+ | x\neq y\}$. For the following, we do not assume that all
rational numbers are represented in a divisor free fraction. The
condition of theorem 1 can be reformulated in the following way:
Iff $p,q,p',q'\in\BZ_+$ such that 
\be\label{eq:det}
  {\rm det}\left(\ba{cc}
    p' & p\\
    q & q'\ea\right) = 1
\ee
then a RCFT exists with $c = 1 - 24pq$ and partition function
$Z[p/q,p'/q'] = (Z[p/q] + Z[p'/q'])/2$ which is unique up to duality. 
Thus, there is a correspondence of points in ${\cal R}(M')$ to elements of
PSL(2,$\BZ$). Furthermore, we see immidiately that the line of unitarity,
$R_1 = R_2$, would correspond to determinant zero matrices.\footnote[2]{
More precisely, only a small subset of them corresponds to $c = 1$
theories. The other determinat
zero matrices correspond to the ordinary $(A_{p-1},A_{q-1})$ minimal
models. For example the matrix ${1\ p\choose q\ pq}$ corresponds to the
partition function $Z = (Z[1/pq]-Z[p/q])/2$. Other types of determinant
zero matrices yield automorphisms of the fusion rules.} Hence, 
non-unitary theories exist arbitrary close to the line of unitarity, but in
order to approximate one rational number by two others keeping condition
(\ref{eq:det}) the denominators will increase and in consequence the
central charge will decrease. Therefore, the degree of non unitarity
increases when approaching the diagonal. Figure 2 shows ${\cal
R}(M')$ calculated up to maximal denominators of $100\,000$.
\begin{figure}[h]
%
  \epsfbox{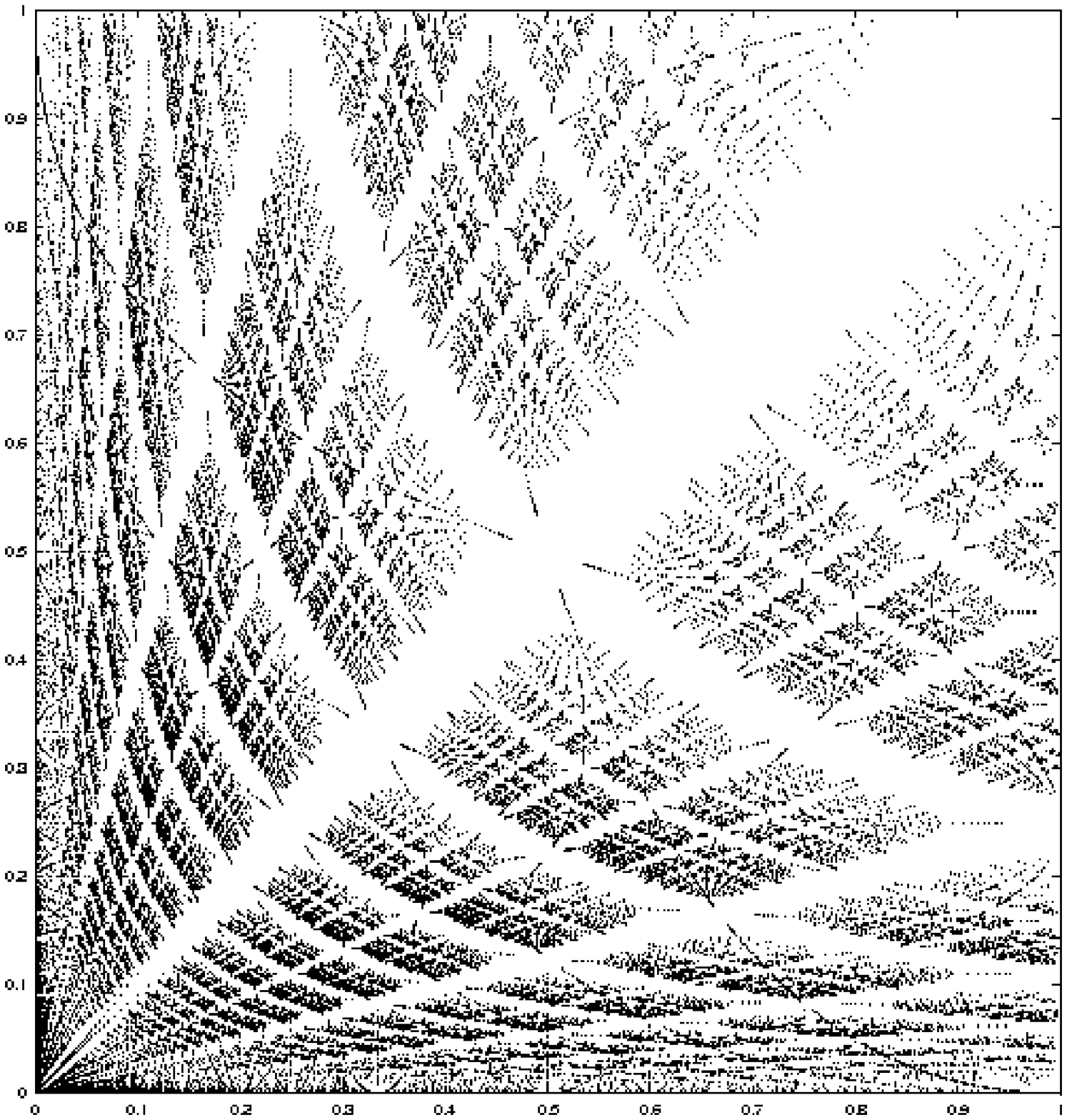}
%
  \medskip\pn
  {\small Fig.\ 2. The moduli space of non unitary $c_{{\em eff}} = 1$
  theories. The $x$-axis is $2R_1^2 = p/q$, the $y$-axis is $2R_2^2 =
  p'/q'$, both in the range $[0,1]$. Shown are points with denominators
  $q,q' \leq 100\,000$. The points outside the shown region are obtained
  by duality $p/q \mapsto q/p$, $p'/q' \mapsto q'/p'$.}
\end{figure}
Note, that in particular near the diagonal the density of points is
rather low. Due to duality we restrict ourselves to the region
$[0,1]\times[0,1]$ in the $2R_1^2,2R_2^2$ plane.
Let us now formulate the following statement:
\begin{claim}
  The set $\Gamma\equiv{\cal R}(M')\cap\{(x,y)\,|\,0\leq y<x\leq 1\}$ is a
  fundamental region of all non unitary RCFTs with $c_{{\it eff}} = 1$.
  ${\cal R}(M')$ is in one-to-one correspondence to the group
  {\rm PSL(2,$\BZ$)} which acts on ${\cal R}(M')$. The correspondence to 
  $\Gamma$ is defined by the following identifications:
  Let $A\in{\rm PSL}(2,\BZ),\ A = {a\ b\choose c\ d}$. Then $A$ is
  identified with $A^T$, $A^{-1}$, and $(A^T)^{-1}$ by duality.
  The symmetry under exchanging $R_1$ with $R_2$ yields the
  identification $A \sim SA$ where $S\equiv{0\ -1\choose 1\ \phantom{-}0}$, 
  i.e.\ we restrict $A$ to matrices with positive integer entries
  only. This subset of {\rm PSL(2,$\BZ$)} forms a semi group which acts on
  $\Gamma$.
\end{claim}
The proof of this statement is simple. $\Gamma$ is obtained by identifying 
theories which are equivalent under duality. Moreover, since the exchange 
of the radii does not affect the partition function, we may order them
such that $0\leq 2R_1^2 < 2R_2^2\leq 1$. The map from the modular group 
PSL(2,$\BZ$) to ${\cal R}(M')$ is just given by 
\be\label{eq:map}
  \left(\ba{cc} 
     a & b\\ 
     c & d
  \ea\right)\mapsto\left(\left|\frac{a}{d}\left|,\vphantom{\frac{b}{d}}
  \right|\frac{b}{c}\right|\right)\,.
\ee
PSL(2,$\BZ$) is completely generated by the two generators
$S\equiv{0\ -1\choose 1\ \phantom{-}0}$ and $T\equiv{1\ 1\choose 0\ 1}$
and the relations $S^2 = (ST)^3 = \Bid$.
Together with (\ref{eq:map}) this yields a very fast algorithm to
generate pictures of ${\cal R}(M')$. The action of $S$ reduces to an
exchange of $R_1$ and $R_2$ which does not affect the partition
function. The set $\Gamma$ is obtained in the following way:
Let $x=|a/d|$, $y=|c/d|$ obtained via (\ref{eq:map}) from any element
of PSL(2,$\BZ$), i.e.\ any admissible word in $S,T$. Then the
corresponding point in $\Gamma$ is given by
$(\max\{\min\{x,1/x\},\min\{y,1/y\}\},\min\{\min\{x,1/x\},\min\{y,1/y\}\})$.
So far, there are always two elements of PSL(2,$\BZ$), which correspond
to one point in ${\cal R}(M')$. They differ by two signs such that,
without loss of generality, ${a\ b\choose c\ d}$ and 
${\phantom{-}a\ -b\choose -c\ \phantom{-}d}$ yield the same point. We
define that matrices with negative integer entries correspond to the
$\BZ_2$-orbifolds (actually, one should multiply all matrix entries
with $2$ to get determinant $4$). This gives the desired one-to-one 
correspondence. 

\mysection{Strange Structure and Quadric Curves}
Figure 2 shows a surprisingly strange and complex structure of the
set ${\cal R}(M')$. First, one observes a ``net of curves''.  We will see 
later that the structure of ${\cal R}(M')$ is also related to continued 
fractions. The self similarity of ${\cal R}(M')$ is
a bit non trivial. The set seems to be divided in rhombi by a net of origin 
lines and symmetric hyperbolas. The self similarity of the rhombi has
two directions. First by rescaling $x\mapsto\alpha x$, $\alpha\in\BQ_+$,
i.e.\ moving along origin lines, and second by moving along the hyperbolas, 
i.e.\ by the mapping $x\mapsto\alpha\frac{1}{x}$. The self similarity is, at
least at the crude approximation level of ${\cal R}(M')$ shown in the figure,
not a perfect one. The ``curves'', which cross through the rhombi,
may differ in shape and number, but the rough structure is the same.
Actually, ${\cal R}(M')$ forms a so called multi-fractal. A multi-fractal
does not have a well defined scaling factor $\alpha$ under which
a part of it reproduces the whole. Instead of that there exist different
scaling factors $\alpha_i$ for which certain subsets are self similar. In our
case every prime number $p$ defines such a scaling and thus ${\cal R}(M')$
is the union of an infinite number of fractals.
\pn
The approximation level may be defined by the maximal length of words in $S,T$,
i.e.\ the number of generators after dividing out the relations 
$S^2 = (ST)^3 = \Bid$, which have been used to generate the picture. 
In the following we will assume that every element
of PSL(2,$\BZ$) is represented by a word of minimal length. Since $T$
acts by translation of either the numerators or the denominators of $x$
and $y$, but $S$ acts by exchanging $x$ and $y$, we see that every
occurence of $S$ in a word $A\in {\rm PSL}(2,\BZ)$ marks a node where
the movement of a point of ${\cal R}(M')$ under the action of $T$
changes the direction. From this the self similarity stems, since this
generates the tree like structere, i.e.\ that the curves seem to spread
out from bundle points. Since the action of $S$ alone does not change
the picture at all, a better measure for the approximation level is just
the maximal number of $T$ generators in a word.
\pn
Let us now discuss these curves. Let $(\frac{p}{q},\frac{p'}{q'})$ be
a point in ${\cal R}(M')$. If $p,q$ are coprime then $a,b\in\BZ$ exist
such that we may write
\be
  \frac{p'}{q'} = \frac{(p')^2}{p'q'} = \frac{(ap+bq)^2}{pq\pm 1}
  =\frac{(ap+bq)^2}{pq}\left(1 \mp \frac{1}{pq\pm 1}\right)\,.
\ee
We see that for $p,q \gg 1$ the deviation from simple algebraic
curves of the form $y=a^2x+2ab+b^2\frac{1}{x}$ gets very small. 
On the other hand every point of ${\cal R}(M')$ lies (more or less) 
close to such a curve. More generally, every point of ${\cal R}(M')$ lies 
close to at least one curve with its graph being of the form 
$(x,\alpha x+\beta\frac{1}{x}+\gamma)$ or
$(\alpha y+\beta\frac{1}{y}+\gamma,y)$ where $\alpha,\beta,\gamma\in\BQ$.
To see this just write
\bea
  \frac{p'}{q'} & = & \frac{(ap+bq)(cp+dq)}{epq+f}\\
                & = & \left(ac\frac{p}{q}+bd\frac{q}{p}+(ad+bc)\right)
                      \left(e +\frac{f}{pq}\right)^{-1}
\eea
or with $p,q \leftrightarrow p',q'$, where $a,b,c,d,e,f\in\BZ$.
Let now $p,q \gg f$. Then there is a high probability that the numerator
divides the denomintor. In particular in the case $a,b$ or $c,d$
coprime there are infinitely many $p,q$ such that $(ap+bq)$ or
$(cp + dq)$ divides $pq + f$ for small $f$. Thus we have proven
\begin{claim}
  Let $(x,y)$ be an arbitrary point of the set ${\cal R}(M')$. Then
  there exist numbers $\alpha,\beta,\gamma\in\BQ$ such that 
  $y\approx \alpha x +\beta\frac{1}{x}+\gamma$ (or vice versa) up to order 
  of at least $\frac{1}{pq}$. In particular there always exists a solution of 
  the form $y\approx a^2x + 2ab + b^2\frac{1}{x}$ if $x$ is given by
  a rational number with coprime numerator and denominator\footnote[3]{
  In this case $a,b\in\BZ$. If the greatest common divisor in the
  rational number $x$ is, say, $m$, then $a,b$ are rational numbers 
  over the modulus $m$.}.
\end{claim}
In the sense of this proposition we may view ${\cal R}(M')$ as the
union of an infinite set of approximated curves, which form the net-like
structure of figure 2. The special curves mentioned in the second part
of the proposition are the ones which collect the points fastest with
increasing approximation depth. Not all parts of the graph of a certain
curve have the same density of collected points. This is due to the
unequally distributed probability that pairs of rational numbers up to 
a maximal denominator lie along a curve.

\mysection{Density in {\it M}'}
The probably most interesting question might be whether the set ${\cal
R}(M')$ of non unitary RCFTs lies dense in the set $M'$. To answer this
question we need some preparation which also enlightens the relationship
of ${\cal R}(M')$, the modular group and continued fractions.
\pn
Consider the mapping (see eqn.\ (\ref{eq:map}))
\bea
  Q:{\rm PSL}(2,\BZ) & \longrightarrow & \BR^2 \\
  A=\left(\ba{cc} a & b \\
                  c & d \ea\right)
                     & \longmapsto     & \left(\frac{a}{d},
                                         \frac{b}{c}\right)\,.
\eea
Define $U_n = (ST^n)^t = {1\ 0\choose n\ 1}{\phantom{-}0\ 1\choose -1\ 0}
= {\phantom{-}0\ 1\choose -1\ n}$.
Then every $A\in{\rm PSL}(2,\BZ)$ can be written as
$A = U_{n_k}\ldots U_{n_2}U_{n_1}(T^{n_0})^t$. We say that $A$ is a
word of length $\ell(A) = \sum_{i=0}^{k}n_i$. It is well known \cite{Hua}
that the fractions $\frac{a}{b}$ and $\frac{c}{d}$ are given as continued
fractions
\bea
  \frac{a}{b} & = & [n_0,n_1,\ldots,n_{k-1}]\,,\\
  \frac{c}{d} & = & [n_0,n_1,\ldots,n_{k-1},n_k]\,,
\eea
where we use the notation
\be
  [n_0,n_1,n_2,\ldots] = n_0 - \frac{1}{{\ds n_1 -
                                \frac{1}{{\ds n_2 - 
                                 \frac{1}{\ddots}}}}}\,.
\ee
Note, that any number $w\in\BR$ has an expansion into a continued
fraction. Now let $A$ be an approximation of a certain number
$w\in\BR$ better than $\eps>0$, i.e.\
$\left|\frac{a}{b}-\frac{c}{d}\right|<\eps$. Then every matrix
$B = A'\cdot A$ such that $\ell(A'\cdot A) = \ell(A')+\ell(A)$ 
yields an approximation of $w$ better then $\eps$. We denote matrices
$B$ with this property by $B\succ A$. Noting that 
$\frac{a/d}{b/c} = \frac{a}{d}\cdot\frac{c}{b}\approx w^2$, 
one then easily sees
\begin{claim}
  The orbit of all matrices $A'\cdot A \succ A$
  under the mapping $Q$ is confined to a band around $w^2$ with width
  $\eps' = 2w\eps + \eps^2$. The band is defined by the equations
  $\frac{x}{y} = \alpha$ or $xy = \alpha$ where $\alpha\in
  [w^2-\eps',w^2+\eps']\cup[\frac{1}{w^2}-\eps',\frac{1}{w^2}+\eps']$.
\end{claim}
Figure 3 shows an example with $A = {\phantom{1}3\ 2\choose 11\ 7}$.
This matrix yields $w^2 \approx \frac{14}{33}$ with an accuracy
$\eps = \left|\frac{2}{3}-\frac{7}{11}\right| = \frac{1}{33}$. The
band is therefore confined to a width $\eps' = 2w\eps + \eps^2 \approx
0.04$. The overall broadness at $x = 1$ is then around $\frac{2}{25}$.
The dirty dust outside this band visible in figure 3 is due to the fact
that the algorithm did not eliminate points with matrices whose length
$\ell(A'\cdot A) < \ell(A')+\ell(A)$.
\begin{figure}[h]
%
  \epsfbox{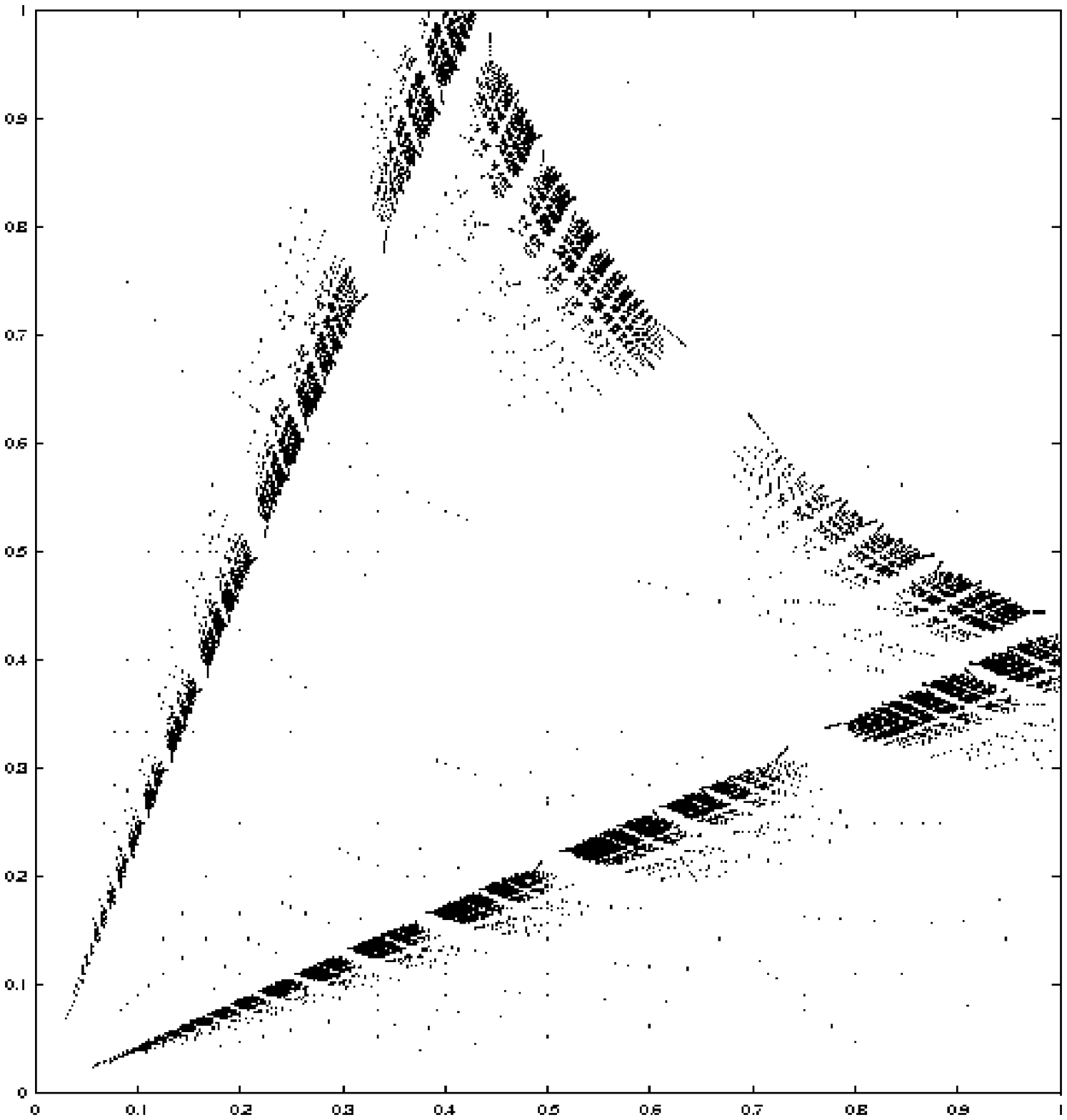}
%
  \medskip\pn
  {\small Fig.\ 3. Orbit of the matrix $A = {\phantom{1}3\ 2\choose 11\ 7}$
  under the action of PSL(2,$\BZ$) using the map $Q$. The orbit is
  confined to a band with width $\eps' \approx 0.04$ around
  $w^2 \approx 0.42$. Plotted are all matrices $A'\cdot A$ with
  $\ell(A')\leq 23$.}
  \end{figure}
\pn
In the limit $\eps\rightarrow 0$ the orbit shrinks to a band with
zero width, since $\eps'\rightarrow 0$. From this it follows that
the band is filled dense by mapping all $B = A'\cdot A \succ A$ to $\BR^2$ via
$Q$, as $\eps\rightarrow 0$ (again $\ell(A'\cdot A) = \ell(A')+\ell(A)$).
In fact, the matrices of the whole modular group PSL(2,$\BZ$)
approximate (via the map $Q$) every real number, if the points are
projected to the, say, $x$-axis. Let now $A$ be given (and thus $w$) and
consider the set of all matrices $B\succ A$. This set forms a closed
semi group. Take the complementary semi group of all matrices $B^{-1}$.
This semi group yields every possible $w'$-values since $A^{-1}$ is
arbitrarily shifted to the left with increasing word length $\ell(B^{-1})$
and thus does not much contribute to the $w'$-value of $B^{-1}$.
Therefore this in some sense complementary set of matrices approximates
every real number under the map $Q$ and projecting onto the $x$-axis. From
our duality relations we then obtain the desired result that 
the original semi group fills the $w$-band dense. We arrive at
\begin{claim}
  The set ${\cal R}(M')$ lies dense in $(\BR_+)^2$, i.e.\ the set of
  non unitary RCFTs with $c_{{\it eff}} = 1$ lies dense in the moduli
  space of all non unitary $c_{{\it eff}} = 1$ CFTs.
\end{claim}
\pn
Finally, for completeness we want to comment on the case of the theories 
with $c = 1 - 24k$, $k$ a half or quarter integer which we did not discuss
here in further detail. These theories have
more complicated partition functions given as certain linear
combinations of (\ref{eq:myzed}). Their partition functions could not
be found by Kiritsis \cite{Kir} since he concentrated on the bosonic
case, i.e.\ required that the $q$-power expansion of the partition
functions had integer spaced exponents. In these fermionic theories
the power series has half integer spaced exponents. For example
the partition functions for the $W(2,3k)$ RCFTs, $k\in\BZ_++\frac{1}{2}$, 
are
\bea
  Z_{{\em ferm}}(\tau,\bar{\tau}) & = &
    Z[2(2l-1)]+Z[2(2l+1)]+Z[\frac{(2l-1)}{2}]+Z[\frac{(2l+1)}{2}]\\
                                  & = &
    Z_{{\em bos}}[2(l-1),2(2l+1)] + 
    Z_{{\em bos}}[\frac{(2l-1)}{2},\frac{(2l+1)}{2}]\,,
\eea
where we have written $k = \frac{2l-1}{2}$ and $Z_{{\em bos}}$ denotes
our bosonic partition function eqn.\ (\ref{eq:myzed}). Of course, one
again may factorize $2l\pm 1$ in different ways to obtain twisted
partition functions, but this has to be done simultaneously in both
bosonic partition functions.
\pn
The set ${\cal R}(M'_{{\em ferm}})$ for these fermionic theories is 
equivalent to our set ${\cal R}(M'_{{\em bos}})$ in the sense that
there exists a bijection given by multiplying one row or column of
a matrix corresponding to a point in ${\cal R}(M'_{{\em bos}})$ by
2. That this indeed is a bijection follows if the several
identifications due to duality and symmetry under exchange of the two
radii are taken into account.
Similar statements hold for the orbifold theories with $k\in\BZ_+ +
\frac{1}{4}$.

\mysection{Conclusion}
We have shown that the structure of the moduli space ${\cal M}$ of
$c_{{\em eff}} = 1$ theories is highly non trivial. The non unitary
RCFTs form a multi-fractal but dense set of isolated points, thus by
no means an algebraic variety or orbifold, as it is conjectured for
the unitary ones (and as it is true for the unitary theories with
$c_{{\em eff}} \leq 1$). This may shed a new light of the possible
structure of the space of all CFTs.
\pn
We proved that the points of the set of non unitary RCFTs are
in one-to-one correspondence with the elements of the modular group
PSL(2,$\BZ$). We constructed an action of PSL(2,$\BZ$) on
${\cal R}(M')$. Moreover, the approximation of the set ${\cal R}(M')$ up
to a certain level (which corresponds to continued fractions up to
a maximal length) yields an interesting structure of quadric curves.
\pn
While these non unitary theories presumably will not have an application in
string theories, they very well may help to describe phenomena in
statistical physics, where non unitarity might be essential. One
example could be the fractional quantum Hall effect (FQHE). One way
to describe the FQHE works with quantum fluid droplets. On the 
border of these droplets lives a CFT which necessarily has $c_{{\em eff}} = 1$.
If the border is not simply connected (e.g.\ the border of an annulus),
then there can be charge transport between different border components.
In this case the CFTs living on the border should be non unitary and 
therefore should be one of the models considered in this work.
\pn
Is has been stated often that only FQHEs to fractional fillings $\nu$ with
small denominators should be observable. In fact, larger denominators
are much more difficult to measure in experiments. This is similar to
our models. Points with larger denominators are always close to points
with rather small denominators. Of course, if a system is forced to go
to one particular theory, it will choose the one with less non
unitarity and a smaller number of states (i.e.\ it will choose the theory
nearest to a trivial ground state theory), and this theory is the one
with smaller denominators. Thus the ``topology'' of the set of our theories 
could very well force a behaviour as it is proposed for the FQHE.
\pn
{\bf Acknowledgement:} We would like to thank A.~Honecker, W.~Nahm, 
M.~R\"osgen and R.~Varnhagen for a lot of discussions, comments and reading 
of the manuscript. We thank the Deutsche Forschungsgemeinschaft for financial 
support.

\end{document}